\documentclass{gapd}

\Type{}
\Title{\centerline{Robust Log-Optimal Strategy with Reinforcement Learning}
}

\Author{\centerline{YiFeng Guo\qquad XingYu Fu\qquad YuYan Shi\qquad MingWen Liu}}{}
\Author{\centerline{Sun Yat-sen University}}{}

\Abstract{We proposed a new Portfolio Management method termed as Robust Log-Optimal Strategy (RLOS), which ameliorates the General Log-Optimal Strategy (GLOS) by approximating the traditional objective function with quadratic Taylor expansion. It avoids GLOS's complex CDF estimation process, hence resists the "Butterfly Effect" caused by estimation error. Besides, RLOS retains GLOS's profitability and the optimization problem involved in RLOS is computationally far more practical compared to GLOS. Further, we combine RLOS with Reinforcement Learning (RL) and propose the so-called Robust Log-Optimal Strategy with Reinforcement Learning (RLOSRL), where the RL agent receives the analyzed results from RLOS and observes the trading environment to make comprehensive investment decisions. The RLOSRL's performance is compared to some traditional strategies on several back tests, where we randomly choose a selection of constituent stocks of the CSI300 index as assets under management and the test results validate its profitability and stability. }

\begin{document}
\maketitle

\section{Introduction}
\label{sec:Introduction}

Portfolio Management (PM), aiming to balance risk and return optimally, sequentially allocating an amount of wealth to a collection of assets in some consecutive trading periods based on investors' return-risk profile [9], is a crucial problem in both theoretical and practical aspects of finance. There are mainly four types of PM strategies [11] proposed from both industry and academia, i.e. Follow-the-Winner, which allocates wealth to the best performed asset; Follow-the-Loser, which allocates wealth to the worst performed asset; Pattern-Matching, which predicts the distribution of price fluctuation of the next trading period based on historical data and chooses the optimal portfolio according to the prediction; Meta-Learning, which combines several different PM strategies to form a better-performed Ensembled investment policy; In our work, we focus on the latter two types, where we ensemble two algorithms from Pattern-Matching category, i.e. Log-Optimal based Strategy and Reinforcement Learning based Strategy, to form our final PM policy.

The General Log-Optimal Strategy (GLOS), originated from information theory [1], attempting to maximize the expectation of the logarithmic rate of return by choosing optimal portfolio weight, is a natural and one of the most renowned PM methods, where many exciting results emerge [10,12,13]. In our work, we will prove rigorously that GLOS is endowed with many elegant characteristics, i.e. Information-Benefit, Greed, and Long-Term Superiority, which justifies its importance. While in practice, the complicated estimation of Cumulative Distribution Function (CDF) and high computational complexity of GLOS make it hard to employ in finance industry, and therefore we propose the so-called Robust Log-Optimal Strategy (RLOS) which avoids GLOS's complex CDF estimation process, hence resists the "Butterfly Effect" caused by estimation error. Besides, RLOS retains GLOS's profitability and the optimization problem involved in RLOS is computationally far more practical compared to GLOS. 

Reinforcement Learning (RL) has witnessed tremendous success in various agent-environment interaction tasks [2,3,4] and it is believed to a possible avenue to the General Artificial Intelligence, where many large companies (e.g. Google DeepMind; Facebook; Baidu etc.) are leading its pioneer research. It is also natural to apply the RL technique to the PM scenario [14,15], where the agent gets reward if its investment decision increases the logarithmic rate of return and gets punishment if it decreases. In our work, we train a RL agent managing a selection of assets on the highly noisy finance environment [5], who receives the analyzed results from RLOS and observes the trading environment using deep Convolutional Neural Network (CNN) simultaneously. The loss function of the RL algorithm, which is composed of reward and penalty, is designed to guide the virtual portfolio manager to trade in a maximizing-profit way. We term the ensemble of RLOS and RL as Robust Log-Optimal Strategy with Reinforcement Learning (RLOSRL).

Our strategies are back-tested on several random selections(Bootstrapping) of constituent stocks of CSI300 index competing with Naïve-Average, Follow-the-Loser, and Follow-the-Winner. RLOSRL and RLOSRL outperforms all these three strategies in all of the back tests, validating the profitability and stability of our algorithm. Further, we can see from the back tests that the trading behaviors of RLOS and RLOSRL are similar to some extent, while RLOSRL defeats RLOS in all back tests, demonstrating the advances of RL.\footnote{The implementation can be viewed at https://github.com/fxy96/Robust-Log-Optimal-Strategy-with-Reinforcement-Learning}

\section{General Log-Optimal Strategy}
\label{sec:Uniqueness}

PM is one of the most common investment methods by managing a collection of assets to optimize a certain objective function. The objective function is often the trade-off between expected risk and expected return for a long-term investment horizon, while in the context of GLOS, the objective function is the expectation of logarithmic rate of return. In this section, we will first define the GLOS formally and then we will study three important characteristics of the strategy, i.e. Information-Benefit, Greed and Long-Term Superiority.

\subsection{Definition of GLOS}
\label{sec:Counteraxeample}

Say we have $d$ assets under management in a single trading period. Let $\textbf{X}=(X_1,X_2,...,X_d)^T$ represent the asset price fluctuation vector in a single trading period and $\textbf{b}=(b_1,b_2,...,b_d)^T$ denote the portfolio weight vector that we set at the start of the trading period. For $\textbf{X}$, each $X_i$ in $\textbf{X}$ is defined as the ratio of the closing price to the opening price of the $i_{th}$ asset over the trading period and $\textbf{X}$ can be deemed as a random vector where we use $\textbf{F}$ to represent its CDF.For $\textbf{b}$, each $b_i$ in $\textbf{b}$ is defined as proportion of the $i_{th}$ asset's investment to the total investment and choosing the optimal $\textbf{b}$ at the start of the trading period is the mission of portfolio manager. 

 In the context of GLOS, the portfolio manager tries to maximize the expected logarithmic rate of return, i.e. choosing $b_X^*$ satisfying the following equation:
 \begin{equation}
\mathbf{b_X^*}\in \mathop{\arg\max}_{\textbf{b}\in B}r_{\textbf{X}}(\textbf{b})
 \end{equation}
 where $r_{\textbf{X}}(\textbf{b})=Elog(\textbf{b}^T\textbf{X})=\int{log(\textbf{b}^T\textbf{x})dF(\textbf{x})}$ is the expected logarithmic rate of return and $B$ is the constrain exerted on $\textbf{b}$.

 The above analysis is set on a single trading period, while in practice, we usually trade on consecutive trading periods, which results in a sequence of $\{\textbf{X}_i\}_{i=1}^n$. If $\{\textbf{X}_i\}_{i=1}^n$ is independent and identically distributed (i.i.d). over different trading periods, then optimal portfolio weight vector which maximizes the expectation of the logarithmic rate of return is the same for every trading period since the objective function is the same and no interdependence of different periods. While in general case, the i.i.d assumption does not hold. 

\subsection{Characteristics of GLOS}
\label{sec:Conclusion}

The reason why we study the GLOS is the elegant mathematical foundations behind it, i.e. Information-Benefit, Greed, and Long-Term Superiority.

\subsubsection{Information-Benefit}

Market Information influences the trading environment directly in a way that can be detected by the GLOS. We call this property as Information-Benefit. 

Say we now trade on a sequence of consecutive trading periods. When new market information arrives at each trading period, the i.i.d assumption of the distribution of  $\{\textbf{X}_i\}_{i=1}^n$ will be violated and therefore we can't just maintain a common portfolio weight vector $b_X^*$ for every trading period. While the information also brings information benefit, i.e. it will increase the expectation of the optimal logarithmic rate of return which will be illustrated below. 

Denote the information by $Y$ and the conditional distribution of $\textbf{X}$ given $Y=y$ by $F(\textbf{X}|Y=y)$ at each trading period. Let $\textbf{b}_{\textbf{X}|Y}^{T*}$ be the optimal portfolio weight vector such that: 
\begin{equation}
\begin{aligned}
\textbf{b}_{\textbf{X}|Y}^{T*}
&\in \mathop{\arg\max}_{\textbf{b}\in B}r_{\textbf{X}|Y}(\textbf{b})\\
&=\mathop{\arg\max}_{\textbf{b}\in B}\int{log(\textbf{b}^T\textbf{x})dF(\textbf{x}|Y=y})
\end{aligned}
\end{equation}

The increment of expected logarithmic rate of return is defined as: 
\begin{equation}
\Delta V_Y=r_{\textbf{X}|Y}(\textbf{b}_{\textbf{X}|Y}^{T*}\textbf{x})-r_{\textbf{X}|Y}(\textbf{b}_{\textbf{X}}^{T*}\textbf{x})
\end{equation}

$\Delta V_Y$ satisfies some elegant mathematical properties which give it some reasonable constrains.
\begin{theorem}
Suppose $\Delta V=E(\Delta V_Y)$. $\Delta V_Y$ and $\Delta V$ satisfies:\\
\begin{enumerate}[1)]
\item $\Delta V_Y \ge 0$
\item $\Delta V_Y \le \int{f_{\textbf{X}|Y=y}(\textbf{x})log\frac{f_{\textbf{X}|Y=y}(\textbf{x})}{f(\textbf{x})}d\textbf{x}}$
\item $\Delta V \le \iint{h(\textbf{x},y)log\frac{h(\textbf{x},y)}{f(\textbf{x})g(y)}d\textbf{x}dy}$ 
\end{enumerate}
where $f,g$ are the marginalized density functions for $\textbf{X}$ and $Y$ respectively and $h$ is the joint density function of  $\textbf{X}$ and $Y$. \\
See the proof in the appendix.
\end{theorem}

Note that $\Delta V_Y$ and $\Delta V$ are controlled by some upper bounds, which means the increment of expected logarithmic rate of return brought by information benefit is not in an unreasonable scale. 

Further, note that $\Delta V$ has an upper bound $\iint{h(\textbf{x},y)log\frac{h(\textbf{x},y)}{f(\textbf{x})g(y)}d\textbf{x}dy}$, which is the mutual information between $\textbf{X}$ and $Y$, a concept frequently studied in information theory [1]. When $\textbf{X}$ and $Y$ are independent, we see that $\Delta V=0$ (Independence implies that the joint density function equals to the product of marginalized density functions), which indicates that no information has been detected by the trading strategy. When $\textbf{X}$ is completely determined by $Y$, this upper bound is exactly the entropy of the information $Y$, another important concept studied in information theory [1]. Further explanation of the upper bound may be intriguing but is out of the scope of this paper and we leave the exploration of this avenue for future research.

\subsubsection{Greed and Long-Term Superiority}

Say we trade on some consecutive trading periods. When $\{\textbf{X}_i\}_{i=1}^n$ is i.i.d, the GLOS maintains a certain fixed portfolio weight vector $b_X^*$ for all trading periods at the start. While, due to the market information which we have discussed in section 2.2.1, the i.i.d. assumption does not hold. Fortunately, the i.i.d. assumption is not necessary. The GLOS can make greedy investment decision for every trading period, i.e. only focusing on maximizing the expected logarithmic rate of return for each single trading period, while the strategy is still superior to other strategies asymptotically in the view of final gross wealth. 

Denote the final gross wealth at the end of the $n_{th}$ trading period with a sequence of portfolio weight vector $\{\textbf{b}_i\}_{i=1}^n$ by $S_n=S_0\Pi_{i=1}^{n}\textbf{b}_i^T\textbf{X}_i$ and the final gross wealth using the GLOS $S_n^*=S_0\Pi_{i=1}^{n}\textbf{b}_i^{*T}\textbf{X}_i$. Next theorem indicates the superiority of the General Log-Optimal Strategy over other PM strategies.

\begin{theorem}
$S_n^*$ is asymptotically superior to $S_n$ with probability 1.\\
\textbf{Proof}: See the proof in the appendix.
\end{theorem}

\section{Robust Log-Optimal Strategy }

There are several disadvantages to implement the GLOS in finance industry. In GLOS, we assume the CDF $F(\textbf{x})$ of price fluctuation vector $\textbf{X}$ is given, while in practice, it's impossible for portfolio manager to know $F(\textbf{x})$ in advance. To implement GLOS, they need to estimate $F(\textbf{x})$ from historical data with certain assumptions, where the estimation error and the improper assumptions may cause the so-called "Butterfly Effect". Besides, even if we have known $F(\textbf{x})$ already, it is computationally expensive to optimize the objective function $\int{log(\textbf{b}^T\textbf{x})dF(\textbf{x})}$ of GLOS since the high dimensionality of $\textbf{X}$ and $\textbf{b}$ and the logarithmic operation in the expression.

Therefore, we propose RLOS, where we don't estimate $F(\textbf{x})$ and maximize $\int{log(\textbf{b}^T\textbf{x})dF(\textbf{x})}$ directly. In RLOS, we introduce a new objective function called the Allocation Utility, which is a quadratic Taylor approximation to the expectation of the logarithmic rate of return, and it is so simple that only involves the expectation and covariance of $\textbf{X}$. Indeed, it can be viewed as the trade-off between the logarithmic return and its squared coefficient of variation. 

RLOS is computationally effective compared to GLOS, and it is robust in the sense that the upper bound of Allocation Utility deviation can be controlled by the $L_1$-norm of portfolio weight vector and $L_{\infty}$-norm of the bias of covariance matrix estimator.  

\subsection{Objective function of RLOS (Allocation Utility) }

The GLOS has (1). However, the optimization problem relies on the distribution function of $\textbf{X}$ ,which is hard to know in practice.

Supposing $\textbf{X}$'s expectation is $\boldsymbol{\mu}$ and its covariance matrix is $\mathbf{\Sigma}$ , in RLOS, we adopt quadratic Taylor expansion to approximate $Elog(\textbf{b}^T\textbf{X})$.

Thus, we define the Allocation Utility, which is the objective function of RLOS, as: 
\begin{equation}
M(\mathbf{b},\boldsymbol{\mu},\boldsymbol{\Sigma})=log(\mathbf{b}^T \boldsymbol{\mu})-\frac{1}{2(\mathbf{b}^T\boldsymbol{\mu})^2}\mathbf{b}^T\mathbf{\Sigma}\mathbf{b}
\end{equation}
$$See\ the\ details\ of\ computation\ in\ the\ appendix.$$

By maximizing $M(\mathbf{b},\boldsymbol{\mu},\boldsymbol{\Sigma})$, which is the mission of RLOS, we can estimate the solution to GLOS in a robust manner. Note that $M(\mathbf{b},\boldsymbol{\mu},\boldsymbol{\Sigma})$ doesn't involve the distribution function of $\textbf{X}$ and the parameters to be estimated are only the expectation and covariance of $\textbf{X}$, which is far more practical than the objective function in GLOS, where the estimation of $F(\textbf{x})$ is required.

Besides, $M(\mathbf{b},\boldsymbol{\mu},\boldsymbol{\Sigma})$ can be deemed as a trade-off between the logarithmic return and its squared coefficient of variation, which somehow coincides with Markowitz's "expected returns – variance of returns" rule [9].

\subsection{Optimal portfolio weight vector for RLOS}

Consider the optimization problem in RLOS:
$$\mathbf{b^{opt}}\in \mathop{\arg\max} M(\mathbf{b},\boldsymbol{\mu},\boldsymbol{\Sigma}), s.t. \mathbf{b}\in B$$

$\mathbf{b^{opt}}$ can be solved analytically for some natural constrain region $𝐵$ if $\boldsymbol{\mu},\boldsymbol{\Sigma}$ are given. For example, we take $B=\{\mathbf{b}|\mathbf{b}^T\mathbf{e}=1,\mathbf{b}^T\boldsymbol{\mu} \ge c, where\ \mathbf{e}=(1,1,...,1)^T\}$. The first constraint follows from the definition of portfolio weight vector and the second constraint accounts for the minimal expected rate of return. We apply Karush-Kuhn-Tucker condition to solve this optimization problem. We provide the computation in the appendix.

From the optimazition procedure, we can find that $\mathbf{b^{opt}}$ depends on the value of $\boldsymbol{\mu}$ and $\boldsymbol{\Sigma}$ . If the estimation of $\boldsymbol{\mu}$,$\boldsymbol{\Sigma}$ deviates from their true values, it may lead to deviation of estimation of optimal portfolio weight, which may result in the so-called "Butterfly Effect". Thus, we need to control our estimation process to reduce estimation deviation. In the next section, we will prove that the RLOS is robust which provides tolerance against reasonable estimation error.

\subsection{Robustness Analysis of RLOS}

Suppose that $\mathbf{\hat{b}^{opt}}$ is the optimal portfolio weight vector estimator by replacing $\boldsymbol{\mu}$ and $\boldsymbol{\Sigma}$ with their estimators: $\hat{\boldsymbol{\mu}}$ and $\hat{\boldsymbol{\Sigma}}$, in the optimization procedure. The estimation error of $\hat{\boldsymbol{\mu}}$ and $\hat{\boldsymbol{\Sigma}}$ may cause the so-called "Butterfly Effect". Hence, we need to study the robustness of RLOS. 

We first give two reasonable assumptions here:

\begin{enumerate}[1)]
\item Suppose $E(\hat{\boldsymbol{\mu}})=\boldsymbol{\mu}$, i.e. $\hat{\boldsymbol{\mu}}$ is $\boldsymbol{\mu}$'s unbiased estimation.\\
\item Let the entry in the $i_{th}$ row and $j_{th}$ column of $\boldsymbol{\Sigma}-\hat{\boldsymbol{\Sigma}}$ be $\sigma_{ij}$. We assume that $\max \limits_{i}\Sigma_{j=1}^n|\sigma_{ij}| \le M$, where $M$ is a positive constant.
\end{enumerate}

From 1), we know $E(\mathbf{\hat{b}}^{\mathbf{opt}^{T}}\boldsymbol{\hat{\mu}})=\mathbf{\hat{b}}^{\mathbf{opt}^{T}}\boldsymbol{\mu}$. By Law of Large Number, $\forall \epsilon >0$ when sample size $n\to \infty$, we have $P(|\mathbf{\hat{b}}^{\mathbf{opt}^{T}}\boldsymbol{\hat{\mu}}-\mathbf{\hat{b}}^{\mathbf{opt}^{T}}\boldsymbol{\mu}|> \epsilon)\to 0$. $\forall \epsilon >0$, $\exists n_0 \in N_+$ such that $\mathbf{\hat{b}}$ satisfies $|\mathbf{\hat{b}}^{\mathbf{opt}^{T}}\boldsymbol{\hat{\mu}}-\mathbf{\hat{b}}^{\mathbf{opt}^{T}}\boldsymbol{\mu}| \le \epsilon$ when sample size is more than $n_0$. Similar analysis with 2) so we don't repeat it here.

Based on the above two assumptions, we now study deviation between $M(\mathbf{\hat{b}^{opt}},,\boldsymbol{\hat{\mu}},\boldsymbol{\hat{\Sigma}})$ and $M(\mathbf{\hat{b}^{opt}},\boldsymbol{\mu},\boldsymbol{\Sigma})$,  which reflects the robustness of RLOS.

\begin{theorem}
Suppose $\mathbf{b}^T \boldsymbol{\mu} \ge c$ and satisfying the assumptions above, the bias of the RLOS following the estimation has an upper bound.
\begin{equation}
\begin{aligned}
&|M(\mathbf{\hat{b}^{opt}},\boldsymbol{\hat{\mu}},\boldsymbol{\hat{\Sigma}})-M(\mathbf{\hat{b}^{opt}},\boldsymbol{\mu},\boldsymbol{\Sigma})| \\
\le &\frac{1}{2c^2}(\sum_{i=1}^n|\hat{b}_i|)^2 \max \limits_{i}\Sigma_{j=1}^n|\sigma_{ij}|
\end{aligned}
\end{equation}\\
\textbf{Proof}: See the proof in the appendix.
\end{theorem}

Thus, if we choose a suitable constant $c_0>0$ and restrict $\sum_{i=1}^n|\hat{b}_i|\le c_0$, we can ensure that the estimation error is controlled. Hence, to achieve a robust optimization result and make $|M(\mathbf{\hat{b}^{opt}},\boldsymbol{\hat{\mu}},\boldsymbol{\hat{\Sigma}})-M(\mathbf{\hat{b}^{opt}},\boldsymbol{\mu},\boldsymbol{\Sigma})|$ controlled by an upper bound, we add an extra $\sum_{i=1}^n|\hat{b}_i|\le c_0$ constraint in the optimization procedure. We can explain the parameter $c_0$ financially, which is larger than 1 if we allow short-selling and equals to 1 if short-selling is forbidden.

\subsection{Implementation of RLOS}

Say we now trade on the $k_{th}$ trading period, to implement RLOS, the portfolio manager need to estimate the parameters involved in the objective function:
\begin{equation}
M(\mathbf{b},\boldsymbol{\mu},\boldsymbol{\Sigma})=log(\mathbf{b}^T \boldsymbol{\mu})-\frac{1}{2(\mathbf{b}^T\boldsymbol{\mu})^2}\mathbf{b}^T\mathbf{\Sigma}\mathbf{b} \tag{4}
\end{equation}

i.e. we need to estimate the expectation $\boldsymbol{\mu}$ and the covariance matrix $\boldsymbol{\Sigma}$ of $\mathbf{X_k}$ in the $k_{th}$ trading period.

To estimate the parameters, our PM strategy selects a collection of trading periods that are similar to the $k_{th}$ trading period and we then calculate $\hat{\boldsymbol{\mu}}$ and $\hat{\boldsymbol{\Sigma}}$ based on the collection, which is the methodology of Pattern-Matching[11]. The problem remains to be solved is how to define similarity between trading periods, where we define it as the Pearson Correlation between market backgrounds of trading periods. We now go specifically into the implementation. 

\subsubsection{Definition of Market Backgrounds}

Consider trading background of the $i_{th}$ trading period, we define it as the price fluctuation matrix reflecting the market situation from $(i-n)_{th}$ to $(i-1)_{th}$ trading periods. Formally, we write it as:
\begin{equation}
Background(i_{th},n)=
\left[\begin{matrix}
x_{1,(i-n)} & \cdots & x_{1,(i-1)}\\
\vdots & \ddots & \vdots\\
x_{(m,1-n)} & \cdots & x_{(m,(i-1))}
\end{matrix} \right] \notag
\end{equation}

where $x_{(a,t)}$ is price fluctuation (ratio of closing price to the opening price) of the $a_{th}$ asset in the $t_{th}$ trading period. Note that we have $n$ as a hyperparameter of background controlling the length of history under consideration for each trading period. To estimate the statistics of $\mathbf{X_i}$ more accurately, in practice, we use multiple $n$ to define multiple backgrounds for the same trading period.

If $Background(i_{th},n)$ and $Background(j_{th},n)$ are "similar" in some sense, then $\mathbf{X_j}$ may containts useful information that can predict the statistics of $\mathbf{X_i}$.

\subsubsection{Definition of Similarity and Similar Trading Periods Selection}

We use Pearson Correlation to define similarity between trading periods. Formally, we write as:
\begin{equation}
\begin{aligned}
&Similar(i_{th},j_{th},n)\\
=&corr^*(Background(i_{th},n),Background(j_{th},n)) \notag
\end{aligned}
\end{equation}
, where $corr^*$ is the Pearson Correlation. We also set hyperparameter $\rho$ as a standard to determine the notion of "similar", where if $Similar(i_{th},j_{th},n) > \rho$, we say $i_{th}$ and $j_{th}$ trading periods are similar and vice versa.

Having defined the notion of "similar", we now proceed to select periods with similar backgrounds with the $k^{th}$ trading period, which forms a set: 
\begin{equation}
S(k,n,\rho)=\{k-n \le i < k | Similar(i_{th},j_{th},n) > \rho\} \notag
\end{equation}

\subsubsection{Algorithm for RLOS}

So far, we have developed enough notions to implement the algorithm for RLOS. Say we trade on the $k_{th}$ period, for different history length $n$, the selection of similar trading periods may vary and therefore we have bunch of estimated parameters, i.e. $\{{\hat{\boldsymbol{\mu}}}_k^{(n)},{\hat{\boldsymbol{\Sigma}}}_k^{(n)}\}^N_{n=1}$. For each $\{{\hat{\boldsymbol{\mu}}}_k^{(n)},{\hat{\boldsymbol{\Sigma}}}_k^{(n)}\}^N_{n=1}$, the optimal portfolio vector $\mathbf{\hat{b}}_{k}^{\mathbf{opt}^{(n)}}$ can be obtained by maximizing the corresponding objective function and we end up with a collection of portfolio vectors and the mission left now is how to ensemble them into a single portfolio vector for trading. Here we assign each $\mathbf{\hat{b}}_{k}^{\mathbf{opt}^{(n)}}$ with a weight $w^{(n)} \in \mathbb{R}$ which reflects the profitability of the vector and then ensemble them linearly.

To summarize the above discussion, we write the algorithm here:
  \begin{algorithm}[htb]  
  \caption{ RLOS}  
  \label{alg:Framwork}  
  \begin{algorithmic}[1]  
    \Require
      $N,\rho,k,Historical Price Data$;  
    \Ensure  
      optimal $\mathbf{\hat{b}}_{k}^{\mathbf{opt}}$ for the $k_{th}\ trading\ period$;  
    \State for $n=2:N$ do  
    \State \quad Construct $S(k,n,\rho)$
    \State \quad if $|S(k,n,\rho)| \le 1$  
    \State \quad \quad continue
    \State \quad else:  
    \State \quad \quad ${\hat{\boldsymbol{\mu}}}_k^{(n)}=Mean(\{X_i|i\in S(k,n,\rho)\})$
    \State \quad \quad${\hat{\boldsymbol{\Sigma}}}_k^{(n)}=Cov(\{X_i|i \in S(k,n,\rho)\})$
    \State \quad \quad$\mathbf{\hat{b}}_{k}^{\mathbf{opt}^{(n)}}=\mathop{\arg\max} M(\mathbf{b},{\hat{\boldsymbol{\mu}}}_k^{(n)},{\hat{\boldsymbol{\Sigma}}}_k^{(n)})$
    \State \quad \quad $w^{(n)}=log(\Pi_{i \in S(k,n,\rho)}\mathbf{\hat{b}}_{k}^{\mathbf{opt}^{(n)}}\mathbf{X_i})$
    \State end for\\
    \Return $\mathbf{\hat{b}}_{k}^{\mathbf{opt}}=(\Sigma w^{(n)}\mathbf{\hat{b}}_{k}^{\mathbf{opt}^{(n)}})/(\Sigma w^{(n)})$
  \end{algorithmic}  
\end{algorithm}  

\section{Robust Log-Optimal Strategy with Reinforcement Learning}

In this paper, we construct an automatic RL trading agent interacts with the low signal-noise ratio stock market environment $E(t)$. Before the opening of the $t_{th}$ trading period of stock market, the agent receives analyzing results $\mathbf{v_t}$ from RLOS and observes the recent historical data $\mathbf{s_t}$ . Then the agent predicts the optimal opening portfolio weights $\mathbf{b_t^{pre}}$ and the logarithmic rate of return of the $t_{th}$ trading period $r_t^{pre}$ based on the information he(she) just receives. If we denote the CNN with parameter $\boldsymbol{\theta}$ as $f_{\boldsymbol{\theta}}$, then we can summarize the above decision-making process as follow:
\begin{equation}
(\mathbf{b_t^{pre}},r_t^{pre})=f_{\boldsymbol{\theta}}(\mathbf{s_t},\mathbf{v_t})
\end{equation}

After predicting the optimal opening portfolio weights $\mathbf{b_t^{pre}}$, we must carry out stock transaction to transform $\mathbf{b_{t-1}^{end}}$ to $\mathbf{b_t^{pre}}$ once the stock market commences. Here we assume that the transaction can be carried out instantly, which implies that we start as $\mathbf{b_t^{pre}}$ at the very beginning of the $t_{th}$ trading period. We also assume that the transaction fee is zero in our work which is a reasonable assumption since the trading frequency of our strategy is relatively low (once a day in the back test) comparing to high frequency trading, where the transaction fee is high, and the corresponding fee is also trivial comparing to the total investment and the stock price fluctuation in our case.

It's common in finance to find non-stationary time series, which means the distribution of financial data may vary as time develops [5]. Therefore, it's crucial that we train our network constantly to fit into the ever-changing environment E(t). To train our network, we define the loss function as follow:
\begin{equation}
\begin{aligned}
L(\boldsymbol{\theta})=&\alpha (r_t^{pre}-r_t^{true})^2- \beta \mathbf{b_t^{opt}}\cdot log(\mathbf{b_t^{pre}})\\
& - \sigma r_t^{true}+ c \vert \vert \boldsymbol{\theta} \vert \vert
\end{aligned}
\end{equation}
where $\alpha, \beta, \sigma, c>0$ are user specified hyperparameters determining the importance of each terms in loss function and the detail of the above loss function will be discussed in section 6.4. SGD algorithm with history reexperience mechanism [4] is used here to minimize the above loss function. 

\subsection{Predicting}

Before the opening of the $t_{th}$ trading period, the RL trader receives two information: $\mathbf{v_t}$ which is the estimated optimal portfolio weights vector predicted by the RLOS algorithm and $\mathbf{s_t}$ which is the recent historical data. We have discussed the RLOS algorithm thoroughly in section 5 and therefore we only discuss the formation of $\mathbf{s_t}$ here.

$\mathbf{s_t}$ is a tensor the reflects the trading circumstances of $d$ assets in recent n trading periods.The trading circumstances that $\mathbf{s_t}$ takes into consideration are the opening, highest, lowest price and the trading volume of each asset in all the n trading periods. Formally, $\mathbf{s_t}$ is a $d\times n \times 4$ tensor where rows represent different assets, columns represent different periods, and the third dimension represents various aspects of the recent trading environment.

The RL trading agent feeds $\mathbf{s_t}$ into the Convolutional Neural Network and plugs $\mathbf{v_t}$ into the feature map tensors before the output layer. The topology of the output layer is a concatenation of a Softmax layer and fully connected layer. It outputs the predicted optimal opening portfolio weights $\mathbf{b_t^{pre}}$ and the predicted logarithmic rate of return of the $t_{th}$ trading period $r_t^{pre}$  respectively.

The structure of the network is shown as Figure1.
\begin{figure} [htp]
\centering
\includegraphics[width=1.77in,height=1.75in]{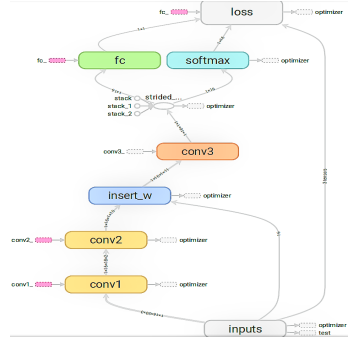}
\caption{Topology of the Network}
\label{fig:graph}
\end{figure}

Filters of the network are all one-dimension for the following two reasons:
\begin{enumerate}[1)]
\item The AI trader can manage different number of stocks without changing the network architecture, and therefore enhance the flexibility of our trading system. 
\item Due to the shared-weight architecture of CNN, the filters are trained to capture some common characteristics of the fluctuation of different stocks independently. 
\end{enumerate}

\subsection{Trading Evaluation}

After setting the opening portfolio weights of the $t_{th}$ trading period to $\mathbf{b_t^{pre}}$, the market commences, and our RL agent do not trade till the opening of next trading period $(t+1)$. At the end of the $t_{th}$ trading period, we can evaluate the performance of the initial portfolio weights $\mathbf{b_t^{pre}}$ and train out neural network later by calculating the logarithmic rate of return $r_t^{true}$.

To recap, we write the formula of again:
\begin{equation}
r_t^{true}=log(\mathbf{b_t^{pre}}\mathbf{X_t})
\end{equation}
where $\mathbf{X_t}$ is the price fluctuation vector that we introduced in section 3.1.    

\subsection{Training}

To recap, we write the loss function here again:
\begin{equation}
\begin{aligned}
L(\boldsymbol{\theta})=&\alpha (r_t^{pre}-r_t^{true})^2- \beta \mathbf{b_t^{opt}}\cdot log(\mathbf{b_t^{pre}})\\
& - \sigma r_t^{true}+ c \vert \vert \boldsymbol{\theta} \vert \vert 
\end{aligned}\tag{7}
\end{equation}
where $\alpha, \beta, \sigma, c>0$ are user specified hyperparameters determining the importance of each terms in loss function. Now we explain each term in detail:

	$\alpha (r_t^{pre}-r_t^{true})^2$:
	this term reflects the squared error between the network predicted logarithmic rate of return and the true logarithmic rate of return, which is calculated at the end of the $t_{th}$ trading period.

	$- \beta \mathbf{b_t^{opt}}\cdot log(\mathbf{b_t^{pre}})$:
	this term reflects the cross entropy between the network predicted optimal opening portfolio weights $\mathbf{b_t^{pre}}$ and the "follow the winner" opening portfolio weights $\mathbf{b_t^{opt}}$ which is calculated by setting $b_{(t,i)}^{opt}=1$ and $b_{(t,j)}^{opt}=0(\forall j\not=i)$ where $X_{(t,i)}$ is largest entry in price fluctuation vector $\mathbf{X_t}$ .The reason why we name $\mathbf{b_t^{opt}}$ as "optimal" portfolio is that:
	$$\mathbf{b_t^{opt}}\in \mathop{\arg\max}log(\mathbf{b}\cdot \mathbf{X_t})$$
	Thus, if $\mathbf{b_t^{pre}}$ is more similar to $\mathbf{b_t^{opt}}$, then we can achieve better logarithmic rate of return $r_t^{true}$.

	$- \sigma r_t^{true}$:
	this term reflects the true logarithmic rate of return in the $t_{th}$ trading period as an instant reward of setting opening portfolio weights to $\mathbf{b_t^{pre}}$. Apparently, the agent tries to maximize $r_t^{true}$, which is the goal of our trading system, by minimizing the loss function $L(\boldsymbol{\theta})$.

	$c \vert \vert \boldsymbol{\theta} \vert \vert $:
	This is a classical term which prevents over-fitting by $L_2$-norm regulation.

After each trading period ends, we need to train our network to update the latest market information. Here we use the history reexperience mechanism[4] by randomly sampling some historical trading records and use SGD algorithm with momentum to minimize the loss function on these records, that is we update the parameter $\boldsymbol{\theta}$ as follow:
\begin{equation}
\theta:=\theta -l \cdot \nabla L(\theta)
\end{equation}
where $l>0$ is the user specified learning rate.Mechanisms like Batch Normalization [6], Optimal Initialization for Relu [8] are used to improve the model's performance.

The purpose of random sampling is to break the correlations of consecutive samples, which is thoroughly studied in [4]. Instead of treating each sample equally as Mnih, V. et al 's work where they use uniform distribution sampling technique, we use Poisson distribution to emphasize the recent samples as the data distribution in finance market is often non-stationary [5]. 

Formally, consider the training at the end of the $t_{th}$ trading period. We sample the $\xi_{th}$ trading record to implement SGD algorithm, where $0<\xi\le t$. In our case, the random variable  $(t-\xi)$ subjects to the Poisson distribution with parameter $\lambda >0$ , that is:
$$(t-\xi)\sim P(\lambda)$$

\section{Back Test}

We use Naïve-Average, Follow-the-Winner and Follow-the-Loser as our baseline PM strategies, competing with the proposed RLOS and RLOSRL in several independent back tests. For each experiment, we randomly select 100 constituent stocks of CSI300 index (use bootstrapping sampling method) as the assets under management and the results of back tests suggest the superiority of our strategies.

\subsection{Back Tests for RLOS}

We run several back tests with different trading length on different stocks to evaluate the RLOS's performance. We set hyperparameters $N$ (the maximal length of historical background for each trading period) to 20 and $\rho$ (the standard for similarity) to 0 in all the experiments. We can see clearly from back tests that RLOS outperforms all the other strategies in both short-term trading and long-term trading. We can see clearly from back tests that RLOS outperforms all the other strategies in both short-term trading and long-term trading. For all the figure below, the horizontal axis represents time and the vertical axis represents total wealth.

\begin{figure}[H]
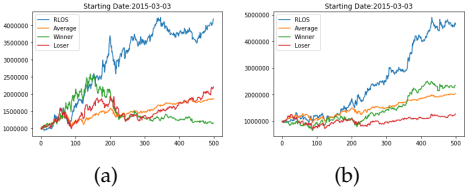

\centering
\subfigure[]{
\includegraphics[width=1.2in]{1.pdf}}
\subfigure[]{
\label{Fig.sub.2}
\includegraphics[width=1.2in]{2.pdf}}
\caption{Trading for 500 days (RLOS) }
\label{Fig.lable}
\end{figure}
\begin{figure}[H]
\centering
\subfigure[]{
\includegraphics[width=1.2in]{3.pdf}}
\subfigure[]{
\label{Fig.sub.2}
\includegraphics[width=1.2in]{4.pdf}}
\caption{Trading for 1000 days (RLOS) }
\label{Fig.lable}
\end{figure}
\begin{figure}[H]
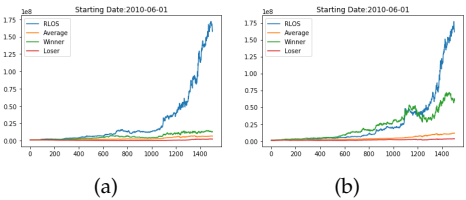

\centering
\subfigure[]{
\includegraphics[width=1.2in]{5.pdf}}
\subfigure[]{
\label{Fig.sub.2}
\includegraphics[width=1.2in]{6.pdf}}
\caption{Trading for 1500 days (RLOS) }
\label{Fig.lable}
\end{figure}

\subsection{Back Tests for RLOSRL}
We validate the performance of RLOSRL on several back tests. The RL agent is trained on data before June $1^{st}$ 2010 to avoid data leakage. To some extent, the trading behaviors of RLOS and RLOSRL are similar since one of the inputs of RLSORL is the analyzed result from RLOS, while we can still see that RLOSRL is superior to RLOS in all the back tests.
\begin{table}
\begin{tabular}{ccc}
\hline
Parameter &Value \\
\hline
$\lambda$: Poisson parameter& 50\\
$r$: Momontum& 0.9\\
$\alpha$& $10^{-4}$\\
$\beta$& $10^{-2}$\\
$\sigma$& $10^{-2}$\\
$c$& $10^{-4}$
\end{tabular}
\caption{Hyperparameter}
\end{table}
\begin{table}
\begin{tabular}{ccc}
\hline
Hundreds of Steps &Learning Rate \\
\hline
0-500& $10^{-2}$\\
0-1000& $10^{-3}$\\
0-1500& $10^{-4}$
\end{tabular}
\caption{Learning Rate Decay}
\end{table}
\begin{figure}[H]
\centering
\subfigure[]{
\includegraphics[width=1.2in]{7.pdf}}
\subfigure[]{
\label{Fig.sub.2}
\includegraphics[width=1.2in]{8.pdf}}
\caption{Trading for 500 days (RLOSRL) }
\label{Fig.lable}
\end{figure}
\begin{figure}[H]
\centering
\subfigure[]{
\includegraphics[width=1.2in]{9.pdf}}
\subfigure[]{
\label{Fig.sub.2}
\includegraphics[width=1.2in]{10.pdf}}
\caption{Trading for 1000 days (RLOSRL) }
\label{Fig.lable}
\end{figure}
\begin{figure}[H]
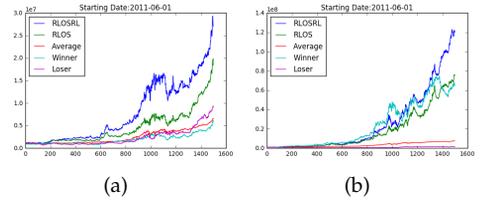

\centering
\subfigure[]{
\includegraphics[width=1.2in]{11.pdf}}
\subfigure[]{
\label{Fig.sub.2}
\includegraphics[width=1.2in]{12.pdf}}
\caption{Trading for 1500 days (RLOSRL) }
\label{Fig.lable}
\end{figure}

\section{Conclusion}

In this paper, we first analyze the advantages of GLOS algorithm, which are Information-Benefit, Greed and Long-Term Superiority, and then propose the so-called RLOS algorithm, which is robust, profitable, and computationally effective, by approximating the objective function of GLOS using Taylor expansion. We further combine the RLOS with RL technique to form an ensembled strategy, where the PM agent is trained in a way to maximize the expected logarithmic rate of return of investment. The stability and profitability of our methods are empirically validated on several independent experiments. We leave some possible avenues for future exploration here:
\begin{itemize}
	\item[$\bullet$] Customize our PM strategies on other types of market. say Future, Currency, Bond etc.
	\item[$\bullet$] Customize our PM strategies on high frequency trading.	
	\item[$\bullet$] Combine assets selection strategy with our PM strategies.
	\item[$\bullet$] Take transaction fee into consideration.
	\item[$\bullet$] Estimate the parameters involved in our model more precisely.
	\item[$\bullet$] Approximate the objective function of GLOS more precisely.
\end{itemize}

\section*{References}

\noindent[1] Cover, T.M. and Thomas, J.A., 2012. Elements of information theory. John Wiley \& Sons.
Vancouver

\noindent[2] Silver, D., Huang, A., Maddison, C.J., Guez, A., Sifre, L., Van Den Driessche, G., Schrittwieser, J., Antonoglou, I., Panneershelvam, V., Lanctot, M. and Dieleman, S., 2016. Mastering the game of Go with deep neural networks and tree search. nature, 529(7587), pp.484-489.

\noindent[3] Silver, D., Schrittwieser, J., Simonyan, K., Antonoglou, I., Huang, A., Guez, A., Hubert, T., Baker, L., Lai, M., Bolton, A. and Chen, Y., 2017. Mastering the game of go without human knowledge. Nature, 550(7676), p.354.

\noindent[4] Mnih, V., Kavukcuoglu, K., Silver, D., Rusu, A.A., Veness, J., Bellemare, M.G., Graves, A., Riedmiller, M., Fidjeland, A.K., Ostrovski, G. and Petersen, S., 2015. Human-level control through deep reinforcement learning. Nature, 518(7540), p.529.

\noindent[5] Lopez de Prado, M., 2018. The 10 Reasons Most Machine Learning Funds Fail.

\noindent[6] Ioffe, S. and Szegedy, C., 2015, June. Batch normalization: Accelerating deep network training by reducing internal covariate shift. In International conference on machine learning (pp. 448-456).

\noindent[7] Nair, V. and Hinton, G.E., 2010. Rectified linear units improve restricted boltzmann machines. In Proceedings of the 27th international conference on machine learning (ICML-10) (pp. 807-814).

\noindent[8] He, K., Zhang, X., Ren, S. and Sun, J., 2015. Delving deep into rectifiers: Surpassing human-level performance on imagenet classification. In Proceedings of the IEEE international conference on computer vision (pp. 1026-1034).

\noindent[9] Markowitz, H., 1952. Portfolio selection. The journal of finance, 7(1), pp.77-91.

\noindent[10] Algoet, P.H. and Cover, T.M., 2011. Asymptotic optimality and asymptotic equipartition properties of log-optimum investment. In THE KELLY CAPITAL GROWTH INVESTMENT CRITERION: THEORY and PRACTICE (pp. 157-179).

\noindent[11] Li, B. and Hoi, S.C., 2014. Online portfolio selection: A survey. ACM Computing Surveys (CSUR), 46(3), p.35.

\noindent[12] Ormos, M. and Urbán, A., 2013. Performance analysis of log-optimal portfolio strategies with transaction costs. Quantitative Finance, 13(10), pp.1587-1597.

\noindent[13] Vajda, I., 2006. Analysis of semi-log-optimal investment strategies. In Prague Stochastics (pp. 719-727).

\noindent[14] Jiang, Z. and Liang, J., 2016. Cryptocurrency portfolio management with deep reinforcement learning. arXiv preprint arXiv:1612.01277.

\noindent[15] Dempster, M.A. and Leemans, V., 2006. An automated FX trading system using adaptive reinforcement learning. Expert Systems with Applications, 30(3), pp.543-552.

\onecolumn
\newpage
\section*{Appendix}
\begin{appendix}
\section{Proof of Theorem 1}

\end{appendix}
\begin{lemma}
$E(log(\phi(x)))\le log(E(\phi(x))),\forall r,v,x \ge 0,\ \phi(x)> 0.$\\
\textbf{Proof}: Since the logarithm function satisfies
$$log(\lambda x_1 + (1-\lambda)x_2)\ge \lambda log(x_1)+(1-\lambda)log(x_2), \lambda \in [0,1]$$
which shows it is a concave function.

Let $\lambda=\frac{x_2-x}{x_2-x_1}$, when $x\in [x_1,x_2]$. Then we have $log(x) \ge \frac{x_2-x}{x_2-x_1}log(x_1)+\frac{x-x_1}{x_2-x_1}log(x_2)$.

The inequality is equivalent to
$$\frac{1}{x-x_1}[log(x)-log(x_1)] \ge \frac{1}{x_2-x_1}[log(x_2)-log(x_1)]$$

Let $x \to x_1$, we have
$$(x_2-x_1)log'(x_1) \ge log(x_2)-log(x_1)$$

Let $x_0=\sum^m_{i=1}{\lambda}_i x_i$. When $\sum^m_{i=1}{\lambda}_i=1, {\lambda}_i >0$.

For each $i$, we have
$${\lambda}_i(x_i-x_0)log'(x_0) \ge {\lambda}_i[log(x_i)-log(x_0)]$$

Thus
$$\sum_{i=1}^m{\lambda}_i(x_i-x_0)log'(x_0) \ge \sum_{i=1}^m{\lambda}_i[log(x_i)-log(x_0)]$$

Since
\begin{equation}
\begin{aligned}
E(log(\phi(x)))&=\int log(\phi(x))dF(x)\\
&=\lim\limits_{t \to \infty } \sum_{k=1}^{\infty}log(\phi(x_k))(F(\frac{k}{n})-F(\frac{k-1}{n}))\\
log(E(\phi(x))&=log(\int \phi(x)dF(x))\\
&=log(\lim\limits_{t \to \infty } \sum_{k=1}^{\infty}\phi(x_k)(F(\frac{k}{n})-F(\frac{k-1}{n}))
\end{aligned}\notag
\end{equation}
and the logarithm function is continuous, let ${\lambda}_k=F(\frac{k}{n})-F(\frac{k-1}{n}), m \to \infty$, then we have
$$\int log(\phi(x))dF(x) \le log(\int \phi(x)dF(x))$$

So $E(log(\phi(x)))\le log(E(\phi(x)))$ as desired.
\end{lemma}

\begin{lemma}
If $b^*$ is the optimal portfolio and $E\frac{b^T X}{b^{*^T}X}$ exists, we have $E\frac{b^T X}{b^{*^T}X} \le 1$, for any other portfolio $b$.\\
\textbf{Proof}: Let $W(b_{\lambda},F)=\int log(b_{\lambda}^T x)dF(x), b_{\lambda}=\lambda b+(1-\lambda)b^*$, where $b$ is another portfolio. When $\lambda=0$, we have $b_0=b^*$. 

According to the definition we have the greatest value
$$W(b_0,F)=W(b^*,F)=\mathop{\max}_{b \in A}\int log(b^T x)dF(x)$$

We say $W(b_0,F) \ge W(b_k,F),\forall k\in[0,1]$ and $\frac{dW(b_{\lambda},F)}{d\lambda}\le 0$ when $\lambda \to 0$ by the definition of derivative.

That is to say,
\begin{equation}
\begin{aligned}
\lim\limits_{\lambda \to 0_+ }\frac{dW(b_{\lambda},F)}{d\lambda}&=\lim\limits_{\lambda \to 0_+ }\frac{1}{\lambda}[W(b_{\lambda},F)-W(b_{0},F)]\\
&=\lim\limits_{\lambda \to 0_+ }\frac{1}{\lambda}[E(log(\lambda b^TX)+(1-\lambda)b^{*^T}X))-E(log(b^{*^T}X))]\\
&=E(\lim\limits_{\lambda \to 0_+ }\frac{1}{\lambda}log(\lambda \frac{b^T X}{b^{*^T}X}+1-\lambda)) \qquad(*)\\
&=E(\lim\limits_{\lambda \to 0_+ }\frac{1}{\lambda}log(1+\lambda (\frac{b^T X}{b^{*^T}X}-1)))\\
&=E(\frac{b^T X}{b^{*^T}X}-1) \qquad(**)\\
&\le 0
\end{aligned}\notag
\end{equation}

The equality (*) can be referred to [5] and the equality (**) is due to the L'Hospital's rule.

\end{lemma}

Then we will give the proof of \textbf{Theorem 1}.
\bigskip

\noindent\it{\textbf{The Proof of Theorem 1:}}
\begin{equation}
\begin{aligned}
\Delta V_Y&=r_{\mathbf{X}|Y}(\mathbf{b}_{\mathbf{X}|Y}^{*^T}\mathbf{x})-r_{\mathbf{X}|Y}(\mathbf{b}_{\mathbf{X}}^{*^T}\mathbf{x})\\
&=\int log(\mathbf{b}_{\mathbf{X}|Y}^{*^T}\mathbf{x})dF(\mathbf{x}|Y=y)-\int log(\mathbf{b}_{\mathbf{X}}^{*^T}\mathbf{x})dF(\mathbf{x}|Y=y)\\
&=\int log\frac{\mathbf{b}_{\mathbf{X}|Y}^{*^T}\mathbf{x}}{\mathbf{b}_{\mathbf{X}}^{*^T}\mathbf{x}}dF(\mathbf{x}|Y=y)\\
&=\int log(\frac{\mathbf{b}_{\mathbf{X}|Y}^{*^T}\mathbf{x}}{\mathbf{b}_{\mathbf{X}}^{*^T}\mathbf{x}}\cdot \frac{f(\mathbf{x})}{f_{\mathbf{x}|Y=y}(\mathbf{x})})dF(\mathbf{x}|Y=y)+\int log\frac{f(\mathbf{x})}{f_{\mathbf{x}|Y=y}(\mathbf{x})}dF(\mathbf{x}|Y=y)\\
&=\int log(\frac{\mathbf{b}_{\mathbf{X}|Y}^{*^T}\mathbf{x}}{\mathbf{b}_{\mathbf{X}}^{*^T}\mathbf{x}}\cdot \frac{f(\mathbf{x})}{f_{\mathbf{x}|Y=y}(\mathbf{x})})dF(\mathbf{x}|Y=y)+\int f_{\mathbf{x}|Y=y}(\mathbf{x})log\frac{f(\mathbf{x})}{f_{\mathbf{x}|Y=y}(\mathbf{x})}d\mathbf{x}\\
&\le log\int \frac{\mathbf{b}_{\mathbf{X}|Y}^{*^T}\mathbf{x}}{\mathbf{b}_{\mathbf{X}}^{*^T}\mathbf{x}}\cdot \frac{f(\mathbf{x})}{f_{\mathbf{x}|Y=y}(\mathbf{x})}dF(\mathbf{x}|Y=y)+ \int f_{\mathbf{x}|Y=y}(\mathbf{x})log\frac{f(\mathbf{x})}{f_{\mathbf{x}|Y=y}(\mathbf{x})}d\mathbf{x} \qquad(lemma 1)\\
&=log\int \frac{\mathbf{b}_{\mathbf{X}|Y}^{*^T}\mathbf{x}}{\mathbf{b}_{\mathbf{X}}^{*^T}\mathbf{x}}dF(\mathbf{x})+\int f_{\mathbf{x}|Y=y}(\mathbf{x})log\frac{f(\mathbf{x})}{f_{\mathbf{x}|Y=y}(\mathbf{x})}d\mathbf{x}\\
&\le log1+\int f_{\mathbf{x}|Y=y}(\mathbf{x})log\frac{f(\mathbf{x})}{f_{\mathbf{x}|Y=y}(\mathbf{x})}d\mathbf{x} \qquad (lemma2)\\
&=\int f_{\mathbf{x}|Y=y}(\mathbf{x})log\frac{f(\mathbf{x})}{f_{\mathbf{x}|Y=y}(\mathbf{x})}d\mathbf{x}
\end{aligned}\notag
\end{equation}

Furthermore, we define $\Delta V=E(\Delta V_Y)$, the expectation of the increment $\Delta V_Y$ with respect to $Y$. Then we will prove $\Delta V$ also has an upper bound. Denote by $G(H)$ the cumulative distribution function of $Y ((X,Y))$, and $g(h)$ the density function of $Y((X,Y))$, we verify that 
\begin{equation}
\begin{aligned}
\Delta V&=\int \Delta V_{Y=y}dG(y)\\
&\le \iint f_{\mathbf{x}|Y=y} log \frac{f_{\mathbf{x}|Y=y}}{f(\mathbf{x})}d\mathbf{x}dG(y)\\
&=\iint f_{\mathbf{x}|Y=y} \cdot g(y)log \frac{f_{\mathbf{x}|Y=y}g(y)}{f(\mathbf{x})g(y)}d\mathbf{x}dy\\
&=\iint h(\mathbf{x},y)log\frac{h(\mathbf{x},y)}{f(\mathbf{x})g(y)}d\mathbf{x}dy
\end{aligned}\notag
\end{equation}

\section{Proof of Theorem 2}

\noindent\it{\textbf{Proof:}}

According to lemma 2, we have $E\frac{S_n}{S_n^*} \le 1$ and
\begin{equation}
\begin{aligned}
Pr(S_n>n^2S_n^*)&=Pr(\frac{S_n}{S_n^*}>n^2)\\
&=\int_{n^2}^{+\infty}dF(\frac{S_n}{S_n^*})\\
& \le \frac{1}{n^2}\int_{n^2}^{+\infty}\frac{S_n}{S_n^*}dF(\frac{S_n}{S_n^*})\\
& \le \frac{1}{n^2}\int_{0}^{+\infty}\frac{S_n}{S_n^*}dF(\frac{S_n}{S_n^*})\\
& \le \frac{1}{n^2} E\frac{S_n}{S_n^*}\\
& \le \frac{1}{n^2}
\end{aligned}\notag
\end{equation}

That is to say,
$$Pr(\frac{1}{n}log\frac{S_n}{S_n^*}>\frac{1}{n}log n^2)\le \frac{1}{n^2}$$
$$\sum_{n=1}^{\infty}Pr(\frac{1}{n}log\frac{S_n}{S_n^*}>\frac{2logn}{n})\le \sum_{n=1}{\infty}\frac{1}{n^2}<\infty$$
and
\begin{equation}
\begin{aligned}
Pr(\overline{\lim\limits_{n \to \infty}}\{\frac{1}{n}log\frac{S_n}{S_n^*}>\frac{2logn}{n})\})&=\lim\limits_{k \to \infty}Pr(\bigcup_{n=k}^{\infty}\{\frac{1}{n}log\frac{S_n}{S_n^*}\})\\
& \le \lim\limits_{k \to \infty} \sum_{n=k}Pr(\{\frac{1}{n}log\frac{S_n}{S_n^*}\})\\
&=0
\end{aligned}\notag
\end{equation}

This implies, $\exists N>0,\forall n>N$, we have
$$\frac{1}{n}log\frac{S_n}{S_n^*} \le \frac{2logn}{n}$$
Thus, we have $\overline{\lim\limits_{n \to \infty}}\frac{1}{n}log\frac{S_n}{S_n^*} \le 0, with\ probability\ 1$ as desired and conclude that $S_n^*$ is asymptotically superior to $S_n$.

\section{Computation of Taylor Expansion}

\begin{equation}
\begin{aligned}
&Elog(\mathbf{b}^T\mathbf{X})\\
\approx& E(logE(\mathbf{b}^T\mathbf{X})+\frac{\mathbf{b}^T\mathbf{X}-E(\mathbf{b}^T\mathbf{X})}{E(\mathbf{b}^T\mathbf{X})}-\frac{(\mathbf{b}^T\mathbf{X}-E(\mathbf{b}^T\mathbf{X}))^2}{2(E(\mathbf{b}^T\mathbf{X}))^2})\\
=&log(\mathbf{b}^T \boldsymbol{\mu})-\frac{1}{2(\mathbf{b}^T\boldsymbol{\mu})^2}\mathbf{b}^T\mathbf{\Sigma}\mathbf{b}
\end{aligned}\tag{4}
\end{equation}

\section{Proof of Theorem 3}
\begin{lemma}
$\forall p,q,q_1,q_2 \in \mathbb{R}$, satisfying $q_1 \le q \le q_2$, we have
$$|p-q| \le \max{p-q,q_2-p}$$
\textbf{Proof}: If $p \ge q, p-q_1 \ge p-q \ge 0$, which refers to $p-q_1 > |p-q|$.

If $p < q, q_2-p \ge q-p > 0$, which refers to $q_2-p > |p-q|$.

we then prove that $|p-q| \le \max{p-q,q_2-p}$.
\end{lemma}

Next, we will study the deviation between $M(\mathbf{\hat{b}^{opt}},\boldsymbol{\hat{\mu}},\boldsymbol{\hat{\Sigma}})$ and $M(\mathbf{\hat{b}^{opt}},\boldsymbol{\mu},\boldsymbol{\Sigma})$.

\begin{equation}
\begin{aligned}
&|M(\mathbf{\hat{b}^{opt}},\boldsymbol{\hat{\mu}},\boldsymbol{\hat{\Sigma}})-M(\mathbf{\hat{b}^{opt}},\boldsymbol{\mu},\boldsymbol{\Sigma})|\\
=&|log(\mathbf{\hat{b}^{{opt}^T}}\boldsymbol{\hat{\mu}})-\frac{1}{2(\mathbf{\hat{b}^{{opt}^T}}\boldsymbol{\hat{\mu}})^2}\mathbf{\hat{b}^{{opt}^T}}\boldsymbol{\hat{\Sigma}}\mathbf{\hat{b}^{opt}}-log(\mathbf{\hat{b}^{{opt}^T}} \boldsymbol{\mu})+\frac{1}{2(\mathbf{\hat{b}^{{opt}^T}}\boldsymbol{\mu})^2}\mathbf{\hat{b}^{{opt}^T}}\boldsymbol{\Sigma}\mathbf{\hat{b}^{opt}}|\\
=&|(log(\mathbf{\hat{b}^{{opt}^T}}\boldsymbol{\hat{\mu}})-log(\mathbf{\hat{b}^{{opt}^T}} \boldsymbol{\mu}))+(\frac{1}{2(\mathbf{\hat{b}^{{opt}^T}}\boldsymbol{\mu})^2}\mathbf{\hat{b}^{{opt}^T}}\boldsymbol{\Sigma}\mathbf{\hat{b}^{opt}}-\frac{1}{2(\mathbf{\hat{b}^{{opt}^T}}\boldsymbol{\hat{\mu}})^2}\mathbf{\hat{b}^{{opt}^T}}\boldsymbol{\hat{\Sigma}}\mathbf{\hat{b}^{opt}})|\\
\le & |log(\mathbf{\hat{b}^{{opt}^T}}\boldsymbol{\hat{\mu}})-log(\mathbf{\hat{b}^{{opt}^T}} \boldsymbol{\mu})|+ |\frac{1}{2(\mathbf{\hat{b}^{{opt}^T}}\boldsymbol{\mu})^2}\mathbf{\hat{b}^{{opt}^T}}\boldsymbol{\Sigma}\mathbf{\hat{b}^{opt}}-\frac{1}{2(\mathbf{\hat{b}^{{opt}^T}}\boldsymbol{\hat{\mu}})^2}\mathbf{\hat{b}^{{opt}^T}}\boldsymbol{\hat{\Sigma}}\mathbf{\hat{b}^{opt}}|\\
=& |log(\frac{\mathbf{\hat{b}^{{opt}^T}}\boldsymbol{\hat{\mu}}}{\mathbf{\hat{b}^{{opt}^T}} \boldsymbol{\mu}})|+ \frac{1}{2}|\frac{1}{(\mathbf{\hat{b}^{{opt}^T}}\boldsymbol{\mu})^2}\mathbf{\hat{b}^{{opt}^T}}\boldsymbol{\Sigma}\mathbf{\hat{b}^{opt}}-\frac{1}{(\mathbf{\hat{b}^{{opt}^T}}\boldsymbol{\hat{\mu}})^2}\mathbf{\hat{b}^{{opt}^T}}\boldsymbol{\hat{\Sigma}}\mathbf{\hat{b}^{opt}}|\\
\le& |log(\frac{\mathbf{\hat{b}^{{opt}^T}}\boldsymbol{\hat{\mu}}}{\mathbf{\hat{b}^{{opt}^T}} \boldsymbol{\mu}})|+ \frac{1}{2}|\frac{1}{(\mathbf{\hat{b}^{{opt}^T}}\boldsymbol{\mu})^2}\mathbf{\hat{b}^{{opt}^T}}\boldsymbol{\Sigma}\mathbf{\hat{b}^{opt}}-\frac{1}{(\mathbf{\hat{b}^{{opt}^T}}\boldsymbol{\hat{\mu}})^2}\mathbf{\hat{b}^{{opt}^T}}\boldsymbol{\hat{\Sigma}}\mathbf{\hat{b}^{opt}}|\\
=&|log(\frac{\mathbf{\hat{b}^{{opt}^T}}\boldsymbol{\hat{\mu}}+\epsilon}{\mathbf{\hat{b}^{{opt}^T}} \boldsymbol{\hat{\mu}}})|+ \frac{1}{2}|\frac{1}{(\mathbf{\hat{b}^{{opt}^T}}\boldsymbol{\mu})^2}\mathbf{\hat{b}^{{opt}^T}}\boldsymbol{\Sigma}\mathbf{\hat{b}^{opt}}-\frac{1}{(\mathbf{\hat{b}^{{opt}^T}}\boldsymbol{\hat{\mu}})^2}\mathbf{\hat{b}^{{opt}^T}}\boldsymbol{\hat{\Sigma}}\mathbf{\hat{b}^{opt}}|\\
=& |log(1+\frac{\epsilon}{\mathbf{\hat{b}^{{opt}^T}} \boldsymbol{\hat{\mu}}})|+ \frac{1}{2}|\frac{1}{(\mathbf{\hat{b}^{{opt}^T}}\boldsymbol{\mu})^2}\mathbf{\hat{b}^{{opt}^T}}\boldsymbol{\Sigma}\mathbf{\hat{b}^{opt}}-\frac{1}{(\mathbf{\hat{b}^{{opt}^T}}\boldsymbol{\hat{\mu}})^2}\mathbf{\hat{b}^{{opt}^T}}\boldsymbol{\hat{\Sigma}}\mathbf{\hat{b}^{opt}}|
\end{aligned}\notag
\end{equation}

Since $\mathbf{\hat{b}^{{opt}^T}} \boldsymbol{\hat{\mu}} \le c$, we have $|log(1+\frac{\epsilon}{\mathbf{\hat{b}^{{opt}^T}} \boldsymbol{\hat{\mu}}})| \le log(1+\frac{\epsilon}{c})$ for the first element of RHS.

For the second element of RHS, according to lemma 3, we have:
\begin{equation}
\begin{aligned}
&|\frac{1}{(\mathbf{\hat{b}^{{opt}^T}}\boldsymbol{\hat{\mu}})^2}\mathbf{\hat{b}^{{opt}^T}}\boldsymbol{\hat{\Sigma}}\mathbf{\hat{b}^{opt}}- \frac{1}{(\mathbf{\hat{b}^{{opt}^T}}\boldsymbol{\hat{\mu}})^2}\mathbf{\hat{b}^{{opt}^T}}\boldsymbol{\Sigma}\mathbf{\hat{b}^{opt}}|\\
\le& \max\{\frac{1}{(\mathbf{\hat{b}^{{opt}^T}}\boldsymbol{\hat{\mu}})^2}\mathbf{\hat{b}^{{opt}^T}}\boldsymbol{\hat{\Sigma}}\mathbf{\hat{b}^{opt}}-\frac{1}{(\mathbf{\hat{b}^{{opt}^T}}\boldsymbol{\hat{\mu}}+\epsilon)^2}\mathbf{\hat{b}^{{opt}^T}}\boldsymbol{\Sigma}\mathbf{\hat{b}^{opt}},\frac{1}{(\mathbf{\hat{b}^{{opt}^T}}\boldsymbol{\hat{\mu}})^2}\mathbf{\hat{b}^{{opt}^T}}\boldsymbol{\hat{\Sigma}}\mathbf{\hat{b}^{opt}}-\frac{1}{(\mathbf{\hat{b}^{{opt}^T}}\boldsymbol{\hat{\mu}}-\epsilon)^2}\mathbf{\hat{b}^{{opt}^T}}\boldsymbol{\Sigma}\mathbf{\hat{b}^{opt}}\}
\end{aligned}\notag
\end{equation}

Since $-\epsilon \le \mathbf{\hat{b}^{{opt}^T}}\boldsymbol{\hat{\mu}}-\mathbf{\hat{b}^{{opt}^T}}\boldsymbol{\mu} \le \epsilon$,
\begin{equation}
\begin{aligned}
&\frac{1}{(\mathbf{\hat{b}^{{opt}^T}}\boldsymbol{\hat{\mu}})^2}\mathbf{\hat{b}^{{opt}^T}}\boldsymbol{\hat{\Sigma}}\mathbf{\hat{b}^{opt}}-\frac{1}{(\mathbf{\hat{b}^{{opt}^T}}\boldsymbol{\hat{\mu}}+\epsilon)^2}\mathbf{\hat{b}^{{opt}^T}}\boldsymbol{\Sigma}\mathbf{\hat{b}^{opt}}\\
=& \frac{1}{(\mathbf{\hat{b}^{{opt}^T}}\boldsymbol{\hat{\mu}})^2}\mathbf{\hat{b}^{{opt}^T}}[\boldsymbol{\hat{\Sigma}}-\frac{\mathbf{\hat{b}^{{opt}^T}}\boldsymbol{\hat{\mu}}}{\mathbf{\hat{b}^{{opt}^T}}\boldsymbol{\hat{\mu}}+\epsilon}\boldsymbol{\Sigma}]\mathbf{\hat{b}^{opt}}\\
\le& \frac{1}{c^2}\mathbf{\hat{b}^{{opt}^T}}[\boldsymbol{\hat{\Sigma}}-\frac{\mathbf{\hat{b}^{{opt}^T}}\boldsymbol{\hat{\mu}}}{\mathbf{\hat{b}^{{opt}^T}}\boldsymbol{\hat{\mu}}+\epsilon}\boldsymbol{\Sigma}]\mathbf{\hat{b}^{opt}}\\
&\frac{1}{(\mathbf{\hat{b}^{{opt}^T}}\boldsymbol{\hat{\mu}})^2}\mathbf{\hat{b}^{{opt}^T}}\boldsymbol{\hat{\Sigma}}\mathbf{\hat{b}^{opt}}-\frac{1}{(\mathbf{\hat{b}^{{opt}^T}}\boldsymbol{\hat{\mu}}-\epsilon)^2}\mathbf{\hat{b}^{{opt}^T}}\boldsymbol{\Sigma}\mathbf{\hat{b}^{opt}}\\
=& \frac{1}{(\mathbf{\hat{b}^{{opt}^T}}\boldsymbol{\hat{\mu}})^2}\mathbf{\hat{b}^{{opt}^T}}[\boldsymbol{\hat{\Sigma}}-\frac{\mathbf{\hat{b}^{{opt}^T}}\boldsymbol{\hat{\mu}}}{\mathbf{\hat{b}^{{opt}^T}}\boldsymbol{\hat{\mu}}-\epsilon}\boldsymbol{\Sigma}]\mathbf{\hat{b}^{opt}}\\
\le& \frac{1}{c^2}\mathbf{\hat{b}^{{opt}^T}}[\boldsymbol{\hat{\Sigma}}-\frac{\mathbf{\hat{b}^{{opt}^T}}\boldsymbol{\hat{\mu}}}{\mathbf{\hat{b}^{{opt}^T}}\boldsymbol{\hat{\mu}}-\epsilon}\boldsymbol{\Sigma}]\mathbf{\hat{b}^{opt}}
\end{aligned}\notag
\end{equation}

The above inequalties immediately imply that
$$|\frac{1}{(\mathbf{\hat{b}^{{opt}^T}}\boldsymbol{\hat{\mu}})^2}\mathbf{\hat{b}^{{opt}^T}}\boldsymbol{\hat{\Sigma}}\mathbf{\hat{b}^{opt}}- \frac{1}{(\mathbf{\hat{b}^{{opt}^T}}\boldsymbol{\hat{\mu}})^2}\mathbf{\hat{b}^{{opt}^T}}\boldsymbol{\Sigma}\mathbf{\hat{b}^{opt}}| \approx \frac{1}{c^2}|\mathbf{\hat{b}^{{opt}^T}}(\boldsymbol{\hat{\Sigma}}-\boldsymbol{\Sigma})\mathbf{\hat{b}^{opt}}|$$

Let $\mathbf{\hat{b}^{{opt}^T}}=(\hat{b_1},\dots \dots,\hat{b_n})^T$, the $i^{th}$  row and $j^th$  column element is $\sigma_{ij}$ of $\Sigma-\hat{\Sigma}$.

We have
\begin{equation}
\begin{aligned}
|\mathbf{\hat{b}^{{opt}^T}}(\boldsymbol{\hat{\Sigma}}-\boldsymbol{\Sigma})\mathbf{\hat{b}^{opt}}|&=|\sum_{i=1}^n \hat{b_i}(\sum_{j=1}^n\hat{b_j}\sigma_{ij})|\\
& \le \sum_{i=1}^n |\hat{b_i}| |\sum_{j=1}^n\hat{b_j}\sigma_{ij}|\\
& \le \sum_{i=1}^n |\hat{b_i}| \sum_{j=1}^n|\hat{b_j}||\sigma_{ij}|\\
& \le \sum_{i=1}^n |\hat{b_i}| (\sum_{j=1}^n|\hat{b_j}|\sum_{i=1}^n|\sigma_{ij}|)\\
& \le \sum_{i=1}^n |\hat{b_i}| \max_i(\sum_{j=1}^n|\hat{b_j}|\sum_{i=1}^n|\sigma_{ij}|)\\
& = (\sum_{i=1}^n |\hat{b_i}|)^2 \max_i \sum_{i=1}^n|\sigma_{ij}|
\end{aligned}\notag
\end{equation}

Combining all the computation above, we obtain
\begin{equation}
\begin{aligned}
&|M(\mathbf{\hat{b}^{opt}},\boldsymbol{\hat{\mu}},\boldsymbol{\hat{\Sigma}})-M(\mathbf{\hat{b}^{opt}},\boldsymbol{\mu},\boldsymbol{\Sigma})|\\
\le &|log(1+\frac{\epsilon}{\mathbf{\hat{b}^{{opt}^T}} \boldsymbol{\hat{\mu}}})|+ \frac{1}{2}|\frac{1}{(\mathbf{\hat{b}^{{opt}^T}}\boldsymbol{\mu})^2}\mathbf{\hat{b}^{{opt}^T}}\boldsymbol{\Sigma}\mathbf{\hat{b}^{opt}}-\frac{1}{(\mathbf{\hat{b}^{{opt}^T}}\boldsymbol{\hat{\mu}})^2}\mathbf{\hat{b}^{{opt}^T}}\boldsymbol{\hat{\Sigma}}\mathbf{\hat{b}^{opt}}|\\
\le & log(1+\frac{\epsilon}{c})+ \frac{1}{2c^2}(\sum_{i=1}^n |\hat{b_i}|)^2 \max_i \sum_{i=1}^n|\sigma_{ij}|\\
= & \frac{1}{2c^2}(\sum_{i=1}^n |\hat{b_i}|)^2 \max_i \sum_{i=1}^n|\sigma_{ij}|
\end{aligned}\notag
\end{equation}

\section{Computation of Optimal Portfolio}

Let $F(\mathbf{b},\alpha, \beta)=log(\mathbf{b}^T \boldsymbol{\mu})-\frac{1}{2(\mathbf{b}^T\boldsymbol{\mu})^2}\mathbf{b}^T\mathbf{\Sigma}\mathbf{b} + \alpha (\mathbf{b^Te}-1)+\beta(c-\mathbf{b^T\mu})$, where $\beta \ge 0$.

Take $\frac{\partial F(\mathbf{b},\alpha, \beta)}{\partial \mathbf{b}}=0$ and consider $\beta(c-\mathbf{b^T\mu})=0, \beta \ge 0, \mathbf{b^Te}-1=0$ simultaneously.
$$\begin{cases}
-\frac{\boldsymbol{\mu}}{\mathbf{b}^T\boldsymbol{\mu}}-\frac{\mathbf{b^Tb}\boldsymbol{\mu \Sigma}}{(\mathbf{b}^T\boldsymbol{\mu})^3}+\frac{\boldsymbol{\Sigma}\mathbf{b}}{(\mathbf{b}^T\boldsymbol{\mu})^2}+\alpha \mathbf{e}-\beta \boldsymbol{\mu}=0& \\
\beta(c-\mathbf{b^T}\boldsymbol{\mu})=0 & \\
\beta \ge 0 & \\
\mathbf{b^Te}-1=0
\end{cases}$$

Multiply $\mathbf{b}^T$ in both sides of first equality, then we have
\begin{equation}
\begin{aligned}
0&=\mathbf{b}^T[-\frac{\boldsymbol{\mu}}{\mathbf{b}^T\boldsymbol{\mu}}-\frac{\mathbf{b^Tb}\boldsymbol{\mu \Sigma}}{(\mathbf{b}^T\boldsymbol{\mu})^3}+\frac{\boldsymbol{\Sigma}\mathbf{b}}{(\mathbf{b}^T\boldsymbol{\mu})^2}+\alpha \mathbf{e}-\beta \boldsymbol{\mu}]\\
&=-1+\alpha -\beta \mathbf{b}^T\boldsymbol{\mu}
\end{aligned}
\end{equation}

(Case 1) If $\beta = 0$, it implies $\alpha = 1$. Therefore, we substitute $\alpha =1, \beta =0$ into equality and get the  value of optimal portfolio $\mathbf{b^{opt}}$ immediately.

(Case 2) If $\beta >0$, we can see$\mathbf{b^T}\boldsymbol{\mu}=c$ by the second equality. According to (10), we have $\alpha=1+\beta c$.
$$\begin{cases}
-\frac{\boldsymbol{\mu}}{\mathbf{b}^T\boldsymbol{\mu}}-\frac{\mathbf{b^Tb}\boldsymbol{\mu \Sigma}}{(\mathbf{b}^T\boldsymbol{\mu})^3}+\frac{\boldsymbol{\Sigma}\mathbf{b}}{(\mathbf{b}^T\boldsymbol{\mu})^2}+\alpha \mathbf{e}-\beta \boldsymbol{\mu}=0& \\
\mathbf{b^T}\boldsymbol{\mu}=c & \\
\alpha=1+\beta & 
\end{cases}$$

After solving equation system above, we can get the  value of optimal portfolio $\mathbf{b^{opt}}$.
\end{document}